# Terahertz Probe of Photoexcited Carrier Dynamics in Dirac Semimetal $Cd_3As_2$

Wei Lu[1], Jiwei Ling[2], Faxian Xiu[2], and Dong Sun[1,3,*]

[1] *International Center for Quantum Materials, School of Physics, Peking University, Beijing 100871, P. R. China*
[2] *State Key Laboratory of Surface Physics and Department of Physics, Fudan University, Shanghai, China*
[3] *Collaborative Innovation Center of Quantum Matter, Beijing 100871, P. R. China*
[*] *Email: sundong@pku.edu.cn*

**Abstract:** The relaxation dynamics of photoexcited quasiparticles of three-dimensional (3D) Dirac semimetals are vital towards their application in high performance electronic and optoelectronic devices. In this work, the relaxation dynamics of photoexcited carriers of 3D Dirac semimetal $Cd_3As_2$ are investigated by transient terahertz spectroscopy. The visible pump-THz probe spectroscopy measurement shows clear biexponential decays with two characteristic time constants. According to the pump-power and temperature dependence, these two characteristic time constants are attributed to the electron phonon coupling (1-4 ps) and anharmonic decay of hot coupled phonons to electronic uncoupled phonons (2-9 ps), respectively. An anomalous electron-optical phonon coupling reduction and a bottleneck slowing of hot optical phonons relaxation are observed with higher excitation intensities similar to that in graphene. On the other hand, the electron-optical phonon coupling can be enhanced due to the phonon frequency broadening and softening at elevated lattice temperature. Furthermore, the transient THz spectrum response is strongly modified by the phonon assisted intraband absorption of hot carriers from a pure electronic Drude model, which is evidenced by a characteristic THz absorption dip in the transient THz absorption spectrum. This absorption dip is pinned by the discrete optical phonon energy that assists the intraband transition enabled by photoexcitation of hot carriers.

**I. Introduction**

Three dimensional (3D) Dirac semimetals has attracted intense research due to both fundamental interest in condensed state physics and their potential practical device application owing to their non-trivial topology related properties such as low-dissipation Dirac transport [1-6], exotic surface state [7], giant magnetoresistance and diamagnetism [6,8-16], and the versatile transitions to interesting Weyl phase by breaking their inversion and time reversal symmetries [14,17-21]. Among many known 3D Dirac semimetals, $Cd_3As_2$ is a very stable compound and own large linear energy-momentum space, which facilitate various experimental studies. On the other hand, the 3D Dirac semimetal, which processes linear energy-momentum dispersion, can be regarded as 3D analogue of graphene, a well-known 2D Dirac semimetal. Although graphene is very well studied before the emergence of 3D Dirac Semimetals and it's under expect that the 3D counterpart should share many common properties due to the similar electronic structure, 3D Dirac semimetals may have some crucial differences with the 2D system originated from (but not limited to) the dimensionality and detail materials properties, such as phonon spectrum, thus it is necessary to revisit these aspects for 3D Dirac semimetals.

Among various properties that are of interest, the (quasi)particle dynamics is crucial for the development of electronic and optoelectronic devices. The interplay and related dynamical processes between carriers, phonons and other quasiparticles have strong influence on the devices performance especially when the device is running at high speed and high field limit. So far, the ultrafast spectroscopy investigations of the (quasi)particle dynamics of $Cd_3As_2$ and other 3D Dirac semimetals are still limited to visible and near/mid-infrared region and the theoretical studies related to these aspects have been limited to hot carriers relaxation [22-27], photoelectric response [28], phonon dynamics and electron-phonon interactions (especially, the acoustic phonon which is dominant at very low temperature in scale of Kelvin where the electronic temperature is lower than the lowest optical phonon branch) [29-32]. The intraband carrier relaxation dynamics, intraband optical response and intraband optical conductivity of 3D Dirac semimetallic $Cd_3As_2$ remains unexplored experimentally, which are crucial for electron transport and optoelectronic operations considering the fact that semimetals are easily doped. Time-resolved optical pump and THz probe spectroscopy is the standard spectroscopic technique in investigating the low photon energy response (1 THz ~ 4meV), especially the intraband carrier relaxation behavior and intraband conductivity of the materials. Moreover, it can also provide the direct experimental parameters for THz related optoelectronic applications such as optical switch working at THz which $Cd_3As_2$ is promised to be applied for [33-35].

Herein, we report a visible pump-THz probe transient spectroscopy study on 3D Dirac semimetal $Cd_3As_2$. The THz measurement is enable by large area of MBE growth of $Cd_3As_2$ [36]. By photoexcitation of hot carriers with visible pump and THz probe of phonon assisted intraband transition of photoexcited carriers, the dynamical intraband relaxation of photoexcited carriers are measured experimentally. The measurement results indicate the hot-electron system interacts intricately with the phonon system and the hot phonon effect plays arole in intraband relaxation of photoexcited carriers. Similar to graphene [37,38], the electron cooling process is dominated initially by electron-optical phonon scattering with a characteristic time constant of few picosecond, which decreases with electron temperature ($T_e$) but increases with lattice temperature ($T_l$) due to the phonon frequency broadening and softening. The hot optical phonons relax into other phonon modes with a lifetime on the order of 10 picosecond which increases with $T_e$ similar to that of graphene [39]. However, in contrary to graphene [40,41], the lifetime increases with $T_l$, implying the up-conversion effect of lower frequency phonons back to high frequency phonons. The other distinct difference from that of graphene is the transient THz spectrum response of $Cd_3As_2$ cannot be fully described by Drude model due to the enhanced phonons scattering enabled by pump excited hot carriers.

## II. Experimental Details

### A. Sample Preparation

Cd3As$_2$ thin films were grown in a Perkin Elmer (Waltham, MA, USA) 425B molecular beam epitaxy system. Fresh cleaved 2-inch mica (thickness is around 70 μm) was used as substrate. The substrate was annealed at 300 °C for 30 min to remove the absorbed molecules at the surface. Then approximate 10 nm of CdTe was deposited as buffer layer to assist Cd$_3$As$_2$ nucleation. After that, Cd$_3$As$_2$ bulk material (99.9999%, American Elements Inc., Los Angeles, CA, USA) was evaporated on to buffer layer at 170 °C. The growth was *in situ* monitored by the reflection high-energy electron diffraction (RHEED) system. Streaky RHEED reveals a flat surface (shown in supplemental material Figure S1a [42]). The thickness of Cd$_3$As$_2$ film is approximate 50 nm.

### B. Sample Characterization

The crystal structure was determined by X-ray diffraction (Bruker D8 Discovery, Bruker Inc., Billerica, MA, USA). A series of {112} diffraction peaks can be identified from the XRD pattern (shown in supplemental material Figure S1b [42]), indicating the growth direction of the thin film.

The thin films were patterned into a 1×2 mm$^2$ hall bar geometry and the magneto-transport measurements were performed in a Physical Property Measurement system (EverCool2, Quantum Design) with magnetic field 0-9 T and temperature 2-300 K (shown in supplemental material Figure S1c-S1f [42]). Within the entire temperature range, the Hall resistance shows linear behavior with negative slope. The sheet carrier density at 2 K is determined to be $1.6 \times 10^{13}$ cm$^{-2}$ and mobility is 6900 cm$^2$V$^{-1}$s$^{-1}$, while $1.85 \times 10^{13}$ cm$^{-2}$ and 5090 cm$^2$V$^{-1}$s$^{-1}$ at 300 K respectively. The Fermi level was determined as 250 meV above the Dirac node from the Shubnikov-de Hass oscillation at low temperature.

## C. Transient THz transmission measurements

To perform terahertz time domain pump probe spectroscopy measurements, a 250 kHz Ti-sapphire amplifier (RegA) system is used to output laser pulses at 800 nm (1.55 eV) with pulse width of 60 fs [43]. The laser was split into three beams: the first beams is either frequency-doubled with a BBO crystal or directly used for the ultrafast pump; the second beam is used to generate THz through a GaAs photoconductive switch; the third beam is used to map out the THz electric field waveform in the time domain through a 0.5 mm thick ZnTe crystal using a standard electro-optic (E-O) sampling technique [44]. The pump beam is mechanically chopped at approximate 1 kHz and the probe THz is detected using a lock-in amplifier referred to as a chopper. The effective bandwidth of the sampling system is ~1.6 THz. In our measurement, the generated THz combined with the optical pump beam are co-linearly polarized and overlapped on the sample with a ~1.2 mm and ~1.6 mm spot-size respectively.

## III. Results and Discussion

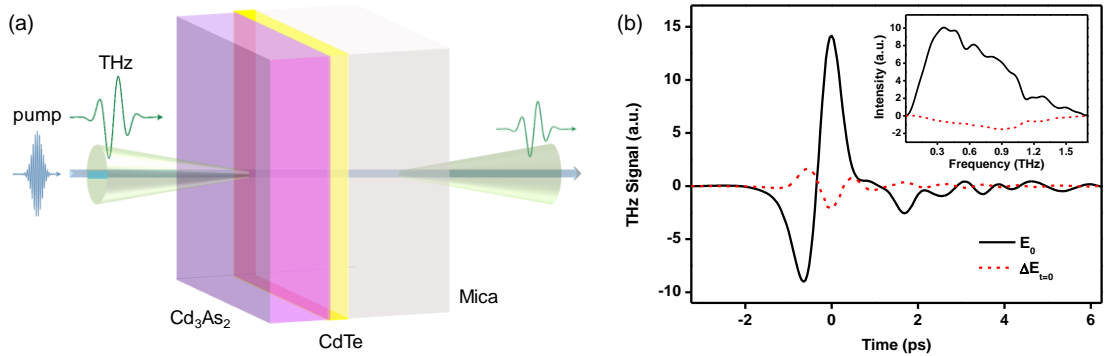

Figure 1. (a) Schematic diagram of sample structure and the transient THz spectroscopy geometry; (b) THz waveform $E_0(t_d)$ (solid line) transmitted through unexcited $Cd_3As_2$ sample and the pump-induced THz transmission changes $\Delta E(t_d)$ at time zero (red dash line). The inset is the corresponding frequency-domain spectrum after Fourier transform of $E_0(t_d)$ and $\Delta E(t_d)$.

As illustrated in the schematic diagram of transient THz spectroscopy measurement (Figure 1a), the THz pulse focused onto the sample co-propagates with the 400-nm (or 800 nm) pump pulse, which excites carriers from the lower (valence like) band to the upper (conduction-like) band across the Dirac point of $Cd_3As_2$. The spot size of the pump is maintained to be larger than the THz probe to ensure a homogeneously photoexcited region is probed by the THz. After rapid carrier-carrier scattering, the photoexcited carriers will thermalize and reach a quasi-equilibrium distribution with elevated electron temperature. Then, the hot carriers will cool down by scattering with phonons [23,25-27,30], which result in dynamical evolution of carrier cooling through electron-phonon scattering. During these processes, the photoexcited hot carriers can enhance the absorption of THz wave compared to unexcited states which is evidenced by the results shown in Figure 1b. At timezero, after photoexcitation, the corresponding pump-induced THz transmission changes $\Delta E(t_d)$ ($t_d$ denotes the delay time of 800-nm sampling pulse in probing the THz wavefrom though E-O sampling), which is obtained by subtracting the transmitted time-domain THz waveform ($E(t_d)$) through the photoexcited $Cd_3As_2$ from that through unexcited one ($E_0(t_d)$), shows clearly inverse sign from $E_0(t_d)$. In the inset of Figure 1b, the corresponding frequency-domain spectrum after Fourier transform of $E_0(t_d)$ and $\Delta E(t_d)$ are plotted, the spectrum shows a maximum around 0.4 THz. The negative value of $\Delta E(t_d)$ further confirms the photoinduced reduction of THz transmission over the whole measured THz spectrum range. The reduction of the THz transmission or the enhancement of THz absorption is due to the increase of intraband absorption induced by the photoexcitation hot carriers. The interband process of THz photon is negligible because the Fermi level of the as-measured sample is 250 meV above the Dirac node according to the transport measurement.

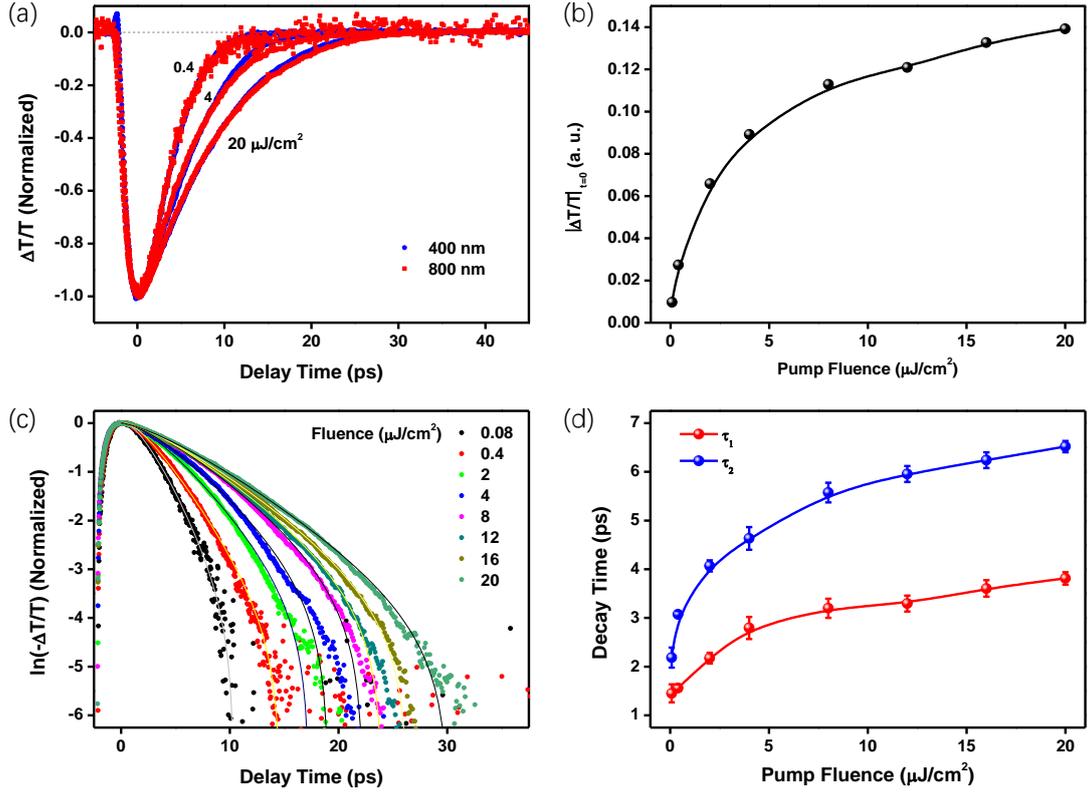

Figure 2. Pump fluence dependence of transient THz dynamics of photoexcited carriers of $Cd_3As_2$ at 78 K. (a) Pump fluences dependence of transient THz dynamics of $Cd_3As_2$ with 400-nm and 800-nm pump excitations respectively. (b) $|dT/T|_{t=0}$ as a function of pump fluence with 400-nm pump, the solid line is guide to the eye. (c) Semi-log plot of the normalized differential transmission response with 400-nm pump excitation and associated fitting curves (solid lines) using the biexponential function: $\Delta T/T = A_0 + A_1 \times \exp(-t/\tau_1) + A_2 \times \exp(-t/\tau_2)$, and (d) the obtained decay time constants $\tau_1$ and $\tau_2$, from biexponential fitting, as a function of pump fluence, the solid lines are guide to the eye.

The transient THz transmission signal, denoted as $\Delta T(t)$, represents the dynamical evolution of pump excited carrier distribution at various pump-probe delay time t. In the transient dynamics measurement, the effect of the buffer layer on the transient THz transmission response is confirmed by recording the response of the buffer layer with the same experimental conditions, where no response from the buffer layer is observed irrespective of pump wavelengths, which indicates the observed response is purely from $Cd_3As_2$.

Figure 2(a) presents the pump fluence dependent transient dynamics of $Cd_3As_2$ with 400- and 800-nm pump excitation measured at 78 K. We note the decay dynamics of transient THz transmission signal of $Cd_3As_2$ are almost independent of the excitation wavelengths. The independence on excitation wavelength is because the timescale of the thermalization (~100 fs) process through electron-electron elastic scattering is much faster than the picosecond time resolution of transient THz measurement. Hereon, we focus our discussion on the results obtained from 400 -nm pump excitation. The magnitude of the transient THz response with 400-nm excitation at time zero ($|dT/T|_{t=0}$), shown in Figure 2b, exhibits clear saturation behavior as the pump fluence increase from 0.04 to 20 $\mu J/cm^2$. This saturation behavior is more salient with 800 nm pump excitation (shown in supplemental material Figure S2a and S2b [42]). Furthermore, according to the normalized natural logarithmic plot of transient response curves (Figure 2c), we notice the relaxation process slows down as the pump fluence increases. More quantitatively, these normalized transient response curves can be fit by bi-exponential function of the form $\Delta T/T = A_0 + A_1 \times \exp(-t/\tau_1) + A_2 \times \exp(-t/\tau_2)$, where $\tau_1$ and $\tau_2$ are the decay time constants and $A_1$ and $A_2$ are the associated amplitudes. As shown in Figure 2d, the decay time constants increase with pump fluence: the fast time constant($\tau_1$) increases from ~1.5 ps to ~ 3.8 ps and the slow constant ($\tau_2$) increases from ~ 2.2 ps to ~6.5 ps as the pump fluence varies from 0.04 to 20 $\mu J/cm^2$.

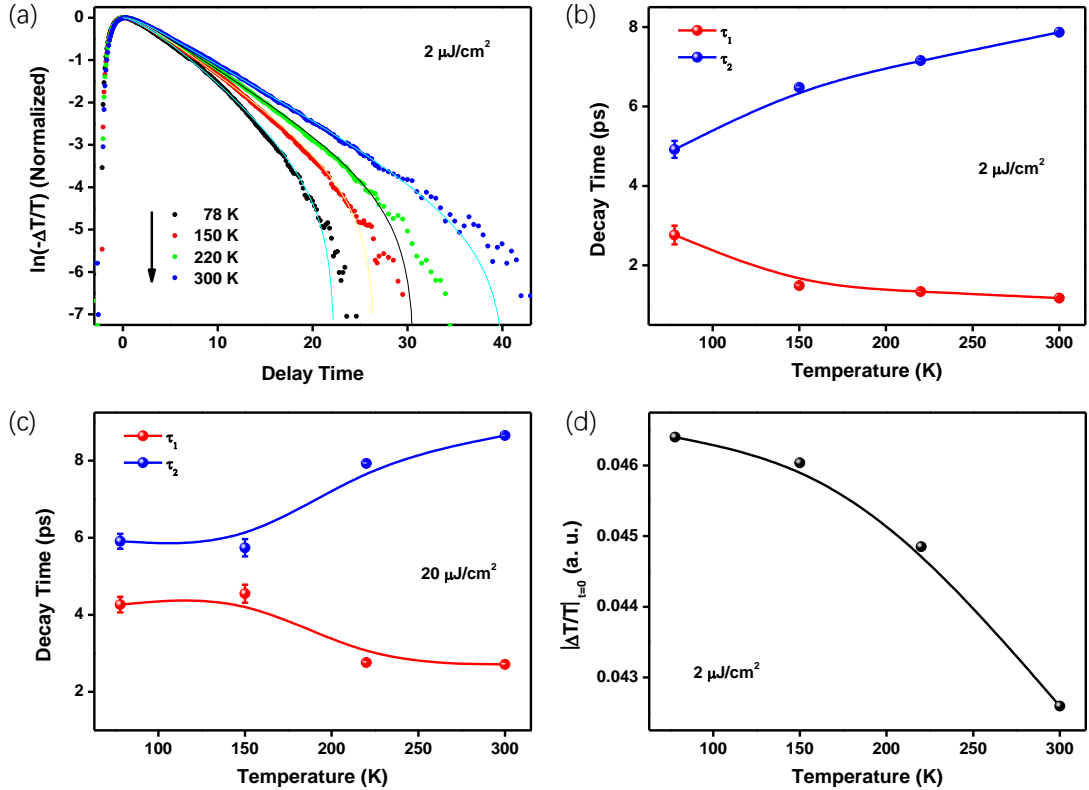

Figure 3. Temperature dependence of transient THz dynamics of $Cd_3As_2$ excited with 400-nm pump. (a) The semi-log plot of normalized transient THz dynamics at different temperatures with pump excitation fluence ~ 2 μJ/cm$^2$. The solid lines are bi-exponential fittings. (b) and (c) are the temperature dependence of fitted decay time constants $\tau_1$ and $\tau_2$ as a function of temperature for fluences of 2 and 20 μJ/cm$^2$, respectively, the solid lines are guide to the eye. (d) Temperature dependent of $|dT/T|_{t=0}$ with excitation of 2 μJ/cm$^2$ pump fluence, the solid line is guide to the eye.

Figure 3 shows the temperature dependent transient THz dynamics measurement results. According to normalized semi-log plot shown in Figure 3a, the decay dynamics gets faster at lower temperature, and this tendency is more obvious with lower pump fluence (Figure S3 in Supplemental Material [42]). According to a biexponential fittings, the slow decay time constant $\tau_2$ increases from ~4.9 ps to ~7.9 ps as temperature changes from 77 K to 300 K for 2 μJ/cm$^2$ pump excitation, which is in stark contrast when compared to temperature dependence of the fast decay time constant $\tau_1$, which decreases from ~2.8 ps to ~1.2 ps (Figure

3b). With 2 μJ/cm$^2$ and 20 μJ/cm$^2$ excitation, both $\tau_1$ and $\tau_2$ change similar amount as the lattice temperature changes from 78K to 300 K (Figure 3b and 3c). Figure 3d shows the $|dT/T|_{t=0}$ also decreases as the temperature increases for fixed pump fluence. The variation in signal intensity is largely originated from the monotonic increase of hot carrier absorption with elevated electron temperature ($T_e$), which will be discussed later.

The transient THz signals manifest the change of $T_e$ as function of pump probe delay (t), and hence the intensity of the signal depends on the difference between electron distribution of excited state at $T_e(t)$ and that of ground state at $T_e=T_l$. Ignoring reflection, the normalized transient THz absorption signals ($\Delta T(t)/T$) can be expressed as $\Delta T(t)/T=[I_0\exp(-\alpha(T_e)d)-I_0\exp(-\alpha(T_l)d)]/[I_0\exp(-\alpha(T_l)d)]=\exp[(\alpha(T_l)-\alpha(T_e))d]-1$, where $I_0$ is the incident THz intensity, $\alpha$ is THz absorption coefficient of the sample, and d is the sample thickness. The intraband transition of THz photon can be induced by hot carriers with a mean energy scale of $2\varepsilon=\hbar\omega+\hbar\omega_q$, where $\omega$ and $\omega_q$ are the frequencies of THz wave and phonon. Neglecting the nonlinear effects in the Fermi golden rule [45-47], the absorption coefficient can be given as:

$$\alpha(T_e) \sim \sum_{k,k',i,j} |M_{k,k',i,j}|^2 [f(E_{k,i},T_e) - f(E_{k',j},T_e)]\delta(E_{k,i} - E_{k',j} + \hbar\omega + \hbar\omega_q) \qquad (1)$$

where, $M_{\mathbf{k},\mathbf{k'},i,j}$ is the transition matrix from the initial state $|\mathbf{k},i\rangle$ with wave vector $\mathbf{k}$ to the final state $|\mathbf{k'},j\rangle$ with the assistant of electron-phonon coupling, $f(E, T_e)$ is the Fermi distribution at $T_e$, $E_{\mathbf{k},i}$ and $E_{\mathbf{k'},j}$ are the energies of initial and final states respectively, and $\delta$ is the delta-function.

According to Eqn. (1), it is straight forward to understand the hot electron induced THz absorption in terms of phonon assisted absorption of THz due to intraband transition of carriers. At elevated electron temperature, the Fermi distribution of hot carriers is broadened over wider energy range, hence, free carriers' absorption of THz undergoes larger possible momentum and energy conservation spaces. When $T_e>T_l$ after photo excitation, $\alpha(T_l)-\alpha(T_e)<0$ due to hot carriers enhanced absorption, thus $\Delta T(t)/T<0$. As $\alpha(T_e)$ increases with $T_e$, the amplitude of $\Delta T(t)/T$ increases with $T_e$. Furthermore, $\Delta T(t)/T$ exhibits saturation behavior as $\alpha(T_e)$ saturates at high electron temperature as evidenced experimentally by the pump power dependence results shown in Figure 2b. On the other hand, the initial electron temperature $T_e(0)$ immediately after photoexcitation with the same pump fluence is almost a constant for measurement at different $T_l$, because the electron heat capacity is relatively small (compared to lattice heat capacity), especially at low $T_e$. As $\alpha(T_l)$ increases with $T_l$, the magnitude of $\Delta T/T|_{t=0}$ decreases as $T_l$ increases (Figure 3d).

The relaxation dynamics of photoexcited hot carriers is schematically shown in Figure 4. The major cooling channel of hot carriers is through phonon scattering. The fast decay time constant ($\tau_1$) is attributed to electron-optical phonon coupling [23-25], while the coupling with acoustic phonon is a lot slower, which should take much longer timescale ranging from nanoseconds to microseconds according to previous studies [29,48]. Another alternative interpretation attributes $\tau_1$ to defects assisted acoustic phonons scattering process which could effectively relax the momentum conservation constraint in electron-phonon coupling [29,48,49] and result in ps decay constant. However, in this diagram, $\tau_1$ should be inversely proportional to both the $T_e(0)$ and $T_l$ [48,49], which contradicts pump fluence dependent results shown in Figure 2. On the other hand, we attribute the slower decay time constant ($\tau_2$) to the anharmonic energy transferring from the hot optical phonons to other uncoupled phonons, according to the previous parallel experimental and theoretical studies on the Dirac materials [24,37,50,51].

The energy transfers from electrons to phonons and from hot optical phonons to other uncoupled phonons can be further described by a set of rate equations based on extended two-temperature model [24,52] as shown below:

$$\frac{dT_e}{dt} = -\beta \frac{T_e - T_{cop}}{\tau_e}$$

$$\frac{dT_{cop}}{dt} = \beta \frac{C_e}{C_{cop}} \frac{T_e - T_{cop}}{\tau_e} - \frac{T_{cop} - T_l}{\tau_{cop}}$$

(2)

where, $T_{cop}$ is the temperature of coupled optical phonons, β is the coefficient that describes the electron-optical phonon coupling strength, $C_e$ and $C_{cop}$ are the specific heats of electron and optical phonon, respectively, and $\tau_e$ and $\tau_{cop}$ are the decay time constant of electron temperature and coupled optical phonon to uncoupled phonon respectively. They are dominated by $\tau_1$ and $\tau_2$ measured in experiment. Here the instantaneous energy transfer process from photon to electron during the pump excitation is ignored, as it takes place at the arrival time of pump pulse with duration of 100-fs which is much shorter than the experimental time resolution that is determined by the duration of THz probe pulse.

The pump fluence dependence results of transient THz dynamics (Figure 2) shows clear signature of electron-phonon coupling reduction for with high pump excitation fluence (with higher transient electronic temperature). This is similar with the pump fluneced dependence measurement results of electron-phonon coupling in other low-dimensional Dirac semimetals such as graphene and metallic nanotubes [45-47,51,53-57], in which phonon stiffenings are observed as result of the reduction of electron-phonon coupling for high transient elctronic temperatures due to the relieving of Kohn anormalies for high energy optical phonons [51,53,55,58]. The observed pump fluence dependence of $\tau_1$ on $Cd_2As_3$ probably share the same origin with graphene and carbon nanotubes. After the quasi-thermal-equilibrium between the hot electron and coupled phonons ($T_e=T_{cop}$), the further cooling of electrons is dominated by the aharmonic phonon coupling process with characteristic decay time constant $\tau_2$. As the pump fluence increases, $\tau_2$ also increases due to the increases of the temperature of uncoupled phonon, which results in phonon bottleneck effect that slows down

$\tau_2$ at high pump fluences [24,59-63]. The same pump fluence dependence of $\tau_2$ is also observed in Dirac semimetal SrMnBi$_2$ [32].

The lattice temperature dependence of fast decay time constant ($\tau_1$) observed in Figure 3c and 3d is an effect of phonon softening. As lattice temperature increases, the atomic bonds become weak and the atoms of Cd$_3$As$_2$ can move more freely, which results in reduction of phonon energy and broadening of the phonon energy spectrum [31]. Therefore, the electron-optical phonon coupling is enhanced due to the relaxation of transition rule, which leads to the increase of $\tau_1$ as temperature decreases. On the other hand, temperature dependence $\tau_2$ indicates decrease of phonon-phonon coupling at higher $T_l$, which is different from that in graphene [40,41]. The energy of optical phonons of Cd$_3$As$_2$ are mostly smaller than 26 meV according to the Raman spectra measurement [31,64,65]. Under the experimental temperature, those low energy optical phonons are mostly already activated, which is not the case in graphene, which has optical phonon energy on the order of 200 meV. With higher experimental temperature, the low energy optical phonons and acoustic phonons has higher probability to convert back to those high energy optical phonons that can couple directly to hot electrons, so $\tau_2$ decreases as temperature decreases which is opposite to the temperature dependence of $\tau_1$. Furthermore, we note that the facilitation of up-conversion of low energy phonons at high temperature [59-61,66] is also consistent with observation of the increment of some low frequency phonons in Raman spectral studies of the Cd$_3$As$_2$ [31,32,65].

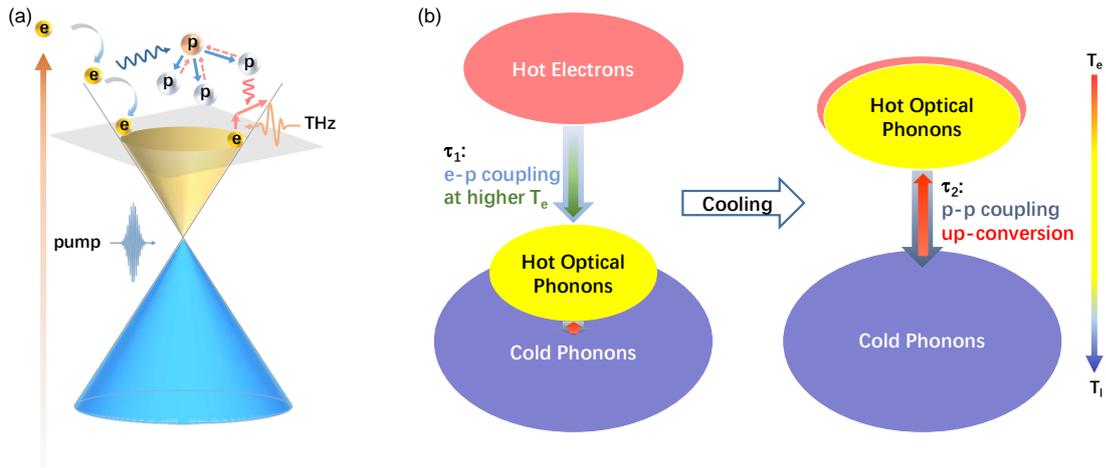

Figure 4. Schematic diagram of the dynamic energy transfers during relaxation of photo excited hot carrier. (a) Energy dispersion of Dirac cones and hot electrons relaxation through emitting optical phonons (e-p coupling), while hot phonons dissipate their energy to cold phonons (p-p coupling). (b) The schematic of two-temperature model that describe the energy transfer during the hot carrier relaxation processes. Hot electrons emit hot optical phonons (e-p coupling) with characteristic time $\tau_1$, while hot phonons dissipate their energy to cold phonons (p-p coupling) with characteristic time $\tau_2$.

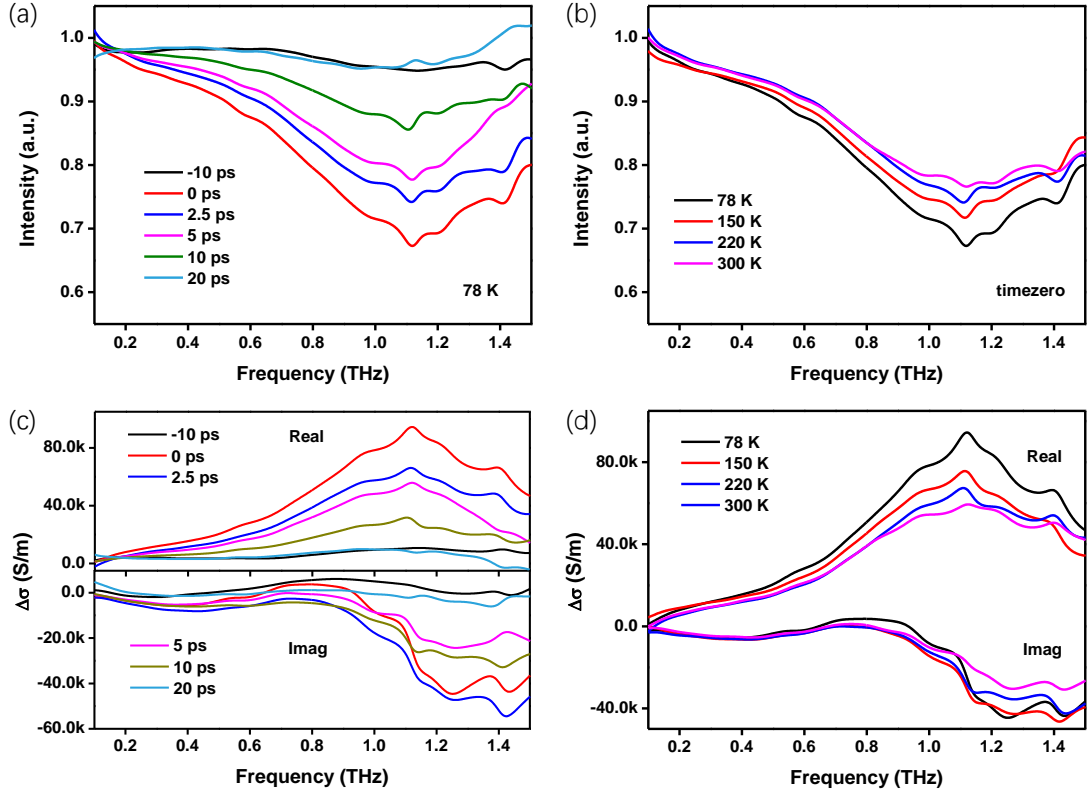

Figure 5. Transient THz spectrum response. Relative THz transmission spectra $E(\omega)/E_0(\omega)$ of $Cd_3As_2$ with 400-nm pump excitation (a) at different delay times at 78 K and (b) at different temperatures at time zero; (c) and (d) the photoinduced changes of THz conductivity $\Delta\sigma(\omega)$ of $Cd_3As_2$ extracted from the pump-induced THz response in (a) and (b), respectively. The pump fluence is 20 μJ/cm² in all measurements.

Figure 5 illustrates the dynamical THz spectrum response after photoexcitation of hot carriers. The transient transmission spectra can provide additional insight regarding the transient change of THz absorption of $Cd_3As_2$ at different pump-probe delays. By applying Fourier transformation to the THz waveform in the time domain, the corresponding un-pumped and pumped frequency-domain complex transmitted THz fields (FDCF) $E_0(\omega)$ and $E(\omega)$ are acquired respectively. Then the photo excited carriers induced relative THz transmissions, are obtained using $E(\omega)/E_0(\omega)$. Figure 5(a) shows the effect of photoexcited hot carriers on the change of THz transmission with 20 μJ/cm² 400-nm excitation for various pump-probe delays at 78 K. The negative dynamical response clearly indicates that the hot carriers induce a significant attenuation of THz transmission over the whole THz frequency range with maximum attenuation of ~30 % around 1.1 THz. This attenuation arises from the strong photoexcited hot carriers' absorption in the $Cd_3As_2$. This negative dynamical response exhibits maximum amplitude at time zero and then it recovers gradually as the pump-probe delay increases. Furthermore, the photoinduced change in the THz transmission also has a clear temperature dependence (Figure 5b): the differential THz transmission signal clearly decreases as temperature increases.

To have more insight on the hot carriers induced transient THz spectrum response of $Cd_3As_2$, we extracted the transient optical conductivity change ($\Delta\sigma(\omega)$) of the $Cd_3As_2$ induced by the pump from the transient transmission signal at fixed pump-probe delay. According to the thin-film approximation, the relation between optical conductivity $\sigma(\omega)$ with FDCF, is given by [38,67]:

$$\sigma(\omega) = \frac{n+1}{Z_0 d}(\frac{E_s(\omega)}{E(\omega)} - 1) \tag{3}$$

where $E_s(\omega)$ is the FDCF of substrate. The corresponding un-pumped and pumped conductivity are $\sigma_0(\omega)$ and $\sigma(\omega)=\sigma_0(\omega)+\Delta\sigma(\omega)$, respectively, and applying the approximation $E_s(\omega) \approx E_0(\omega)$, $\Delta\sigma(\omega)$ can be write as:

$$\Delta\sigma(\omega) = \sigma(\omega) - \sigma_0(\omega) \approx \frac{n+1}{Z_0 d}(\frac{E_0(\omega)}{E(\omega)} - 1) \tag{4}$$

where, we use n=2.73 at 1 THz for the refractive index of the mica substrate [68], which is nearly dispersion free, $Z_0 = 377$ Ω is the impedance of free space, Although the sample contains a very thin (10 nm) CdTe buffer layer, parallel pump-probe experiment on the bear buffer layer on substrate shows negligible transient transmission signal, thus the influence of buffer layer can be safely ignored in the analysis. In case of samples containing only one thin conductive film and one insulating substrate, equation (4) is valid and it serves as good approximation for our experimental scheme.

Figure 5c and 5d show the calculated $\Delta\sigma(\omega)$ of the $Cd_3As_2$ for different pump-probe delays and experimental temperatures. The real and imaginary parts of $\Delta\sigma$ are purely positive and negative, respectively. Both parts gradually recover as pump-probe delay increases (Figure 5c). The major feature of the transient THz transmission spectra of the real part of $\Delta\sigma$ is a wide peak center at 1.1 THz. Corresponding to this absorption peak in the real part, the imaginary part of $\Delta\sigma$ exhibits a sharp slope over the same frequency range, which is consistent according to the Kramer-Kronig relationship. The hot carrier induced THz absorption in a metal/semimetal, taking graphene as an example [38,69,70], can usually be fitted very well with Drude model. In standard Drude model $\sigma=\sigma_{dc}/(1-i\omega\tau)$ [67,71], where $\tau$ is the mean scattering time of the carrier and $\sigma_{dc}=ne\mu=ne^2\tau/m^*$ is the steady-state conductivity, $\Delta\sigma$ can be calculated by subtracting the unexcited Drude conductivity from the excited one. However, all attempts to fit the $\Delta\sigma$ utilizing Drude model fail (supplemental material Figure S4 [42]). This indicates the transient absorption peak center at 1.1 THz is not due to pure hot carrier absorption.

Instead, low energy phonon assisted hot carriers absorption of THz may modulate the Drude response and account for the observed absorption peak center at 1.1 THz. This is because THz absorption of free carriers should fulfil energy and momentum conservation with assistance of certain phonons as implied by Eqn. 1. The excitation of hot carriers can activate the additional phonon mode to participate in the THz absorption that are otherwise forbidden without photo excitation of hot carriers [72-75]. The studies of the phonon spectrum of $Cd_3As_2$ show that there are many low frequency optical phonon modes, such as $A_{1g}$, $B_{1g}$, $B_{2g}$ and $E_g$, may account for this phonon assisted hot carrier absorption effect [31,32,64,65]. As the pump-probe delay time changes, it is evident from Figure 5a and 5c that the transient absorption peak stays at the same frequency which is pinned by related optical phonon energy that is newly activated by hot carriers. As the temperature increases, the central absorption peak frequency has a red shift (Figure 5b), this is consistent with temperature dependent phonon softening behavior. On the other hand, we note depth of the absorption dip increases as temperature decreases as shown in Figure 5b and 5d. This is because that comparing to high lattice temperature, the excitation of hot carriers activate more phonon assisted THz absorption change at low lattice temperature, which induces deeper transient absorption peaks, so this is also consistent with the interpretation of phonon assisted hot carrier induced absorption of THz.

## IV. Conclusions

In summary, we have studied the ultrafast photoexcited quasiparticle relaxation dynamics in $Cd_3As_2$ using transient THz spectroscopy. Through the excitation power and lattice temperature dependent measurement, we can identify the initial fast relaxation of hot carriers is mainly due to the electron-optical phonon scattering and the subsequent relatively slower relaxation is mediated by anharmonic decay of hot phonon into other phonon modes. For the fast decay component, an anomalous reduction of electron-optical phonon coupling is observed at elevated electron temperature under higher pump excitation, similar to that in graphene. However, with elevated lattice temperature, the electron-optical phonon coupling can be enhanced due to the phonon frequency broadening and softening. Furthermore, the presence of hot electrons induces a THz absorption enhancement with a wide peak centered at 1.1 THz due to the modification of phonon assisted intraband absorption by the photoexcitation of hot carriers. This modification makes the transient THz spectrum response different from a pure Drude model. The fast transient dynamics observed in this work implies $Cd_3As_2$ is a promising material platform for ultrafast optoelectronic component such as optical switch working in THz frequency [26].

**Acknowledgements**

This project has been supported by the National Basic Research Program of China (973 Grant No. 2014CB920900), the National Natural Science Foundation of China (NSFC Grant Nos. 11674013, 11704012, 91750109, 11474058 and 61674040), National Key Research and Development Program of China (Grant Nos: 2016YFA0300802, 2017YFA0303302 and 2018YFA0305601), the Recruitment Program of Global Experts, the Open Fund of State Key Laboratory of Precision Measurement Technology and Instruments and the China Postdoctoral Science Foundation (Grant No:2017M610009).

# Supplemental Material for "Terahertz Probe of Photoexcited Carrier Dynamics in Dirac Semimetal Cd₃As₂"

**S1. Structure and Transport Characterizations of Cd$_3$As$_2$ film.**

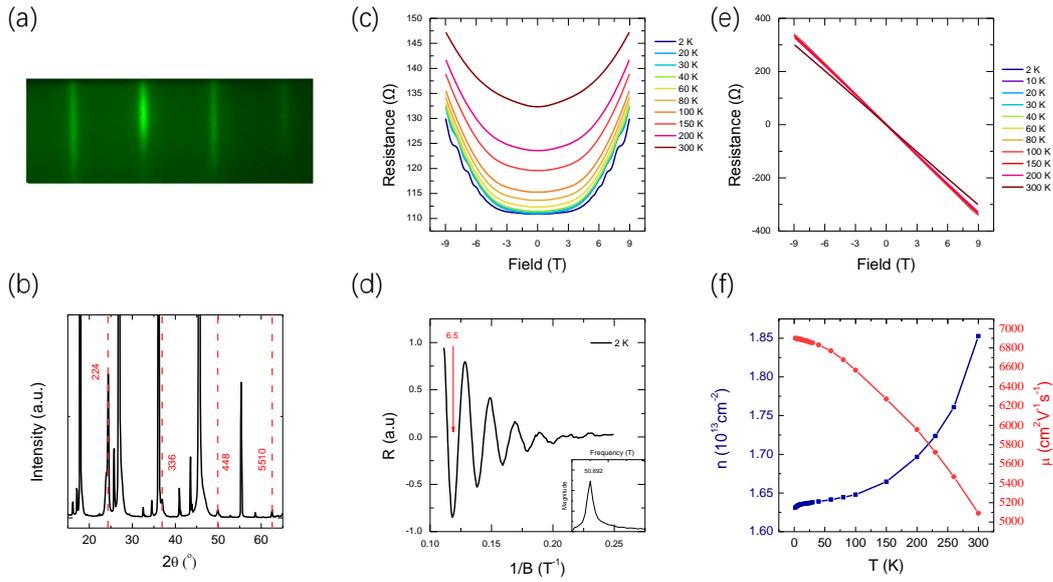

Figure S1. (a) Reflection high-energy electron diffraction pattern, (b) X-ray diffraction spectrum, (c) magnetoresistance, (d) Shubnikov-de Hass oscillation at 2K, insert: the FFT spectrum: the frequency of the oscillation is around 50T, corresponding to $E_F$ = 250meV, (e) Hall resistance: Hall data is linear in the whole temperature range, and (f) temperature dependencies of mobility and carrier density.

**S2. THz absorption saturation and pump fluence dependent carriers relaxation dynamics of Cd$_3$As$_2$ with 800 nm pump**

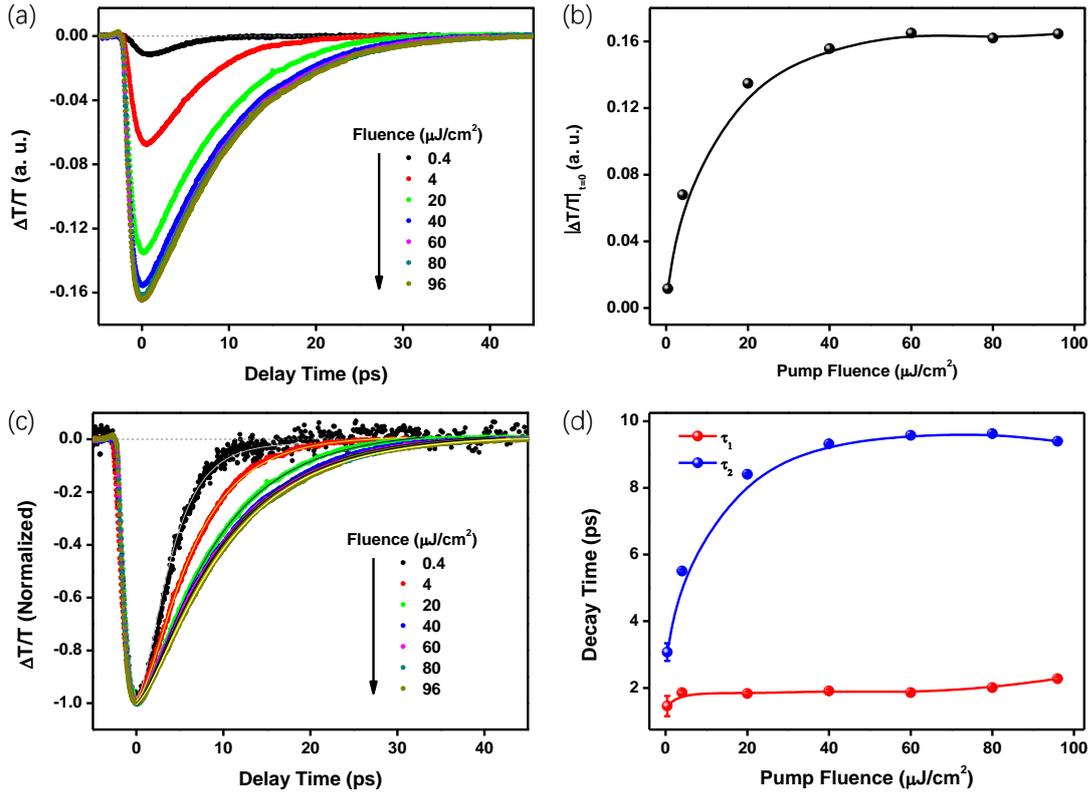

Figure S2. Pump fluence dependence of transient THz dynamics of photoexcited carriers of $Cd_3As_2$ with 800 nm pump at 78 K. (a) Pump fluences dependence of transient THz responses of $Cd_3As_2$, (b) $|dT/T|_{t=0}$ as a function of pump fluence, the solid line is guide to the eye. (c) the normalized differential transmission responses and associated fitting curves (solid lines) using the biexponential function: $\Delta T/T = A_0 + A_1 \times \exp(-t/\tau_1) + A_2 \times \exp(-t/\tau_2)$, and (d) the fitted decay time constants $\tau_1$ and $\tau_2$ as a function of pump fluence, the solid lines are guide to the eye.

Figure S2(a) shows the negative response of $\Delta T(t)/T$ with 800 nm pump excitation. The magnitude of the transient THz responses at time zero ($|dT/T|_{t=0}$) as fluence varies from 0.4 to 96 μJ/cm² are plotted in Figure S2(b) and is apparent that the signal intensity saturates as the fluence becomes larger than 40 μJ/cm², because of the saturation of absorption coefficient at high electron temperature. Correspondingly, the carrier relaxation slows down and saturates at higher pump fluence, as shown in Figure S2(c). The relaxation curves are well fitted by a bi-exponential function, the fast decay time constant $\tau_1$ increases from ~1.5 ps to ~2.3 ps and the slow decay time constant $\tau_2$ increase from ~3.1 ps to ~9.6 ps (Figure S2(d)) with pump fluence increasing.

**S3. Temperature dependent transient response and carriers relaxation dynamics of $Cd_3As_2$ with 400 nm pump excitation and 20 $\mu J/cm^2$ fluence.**

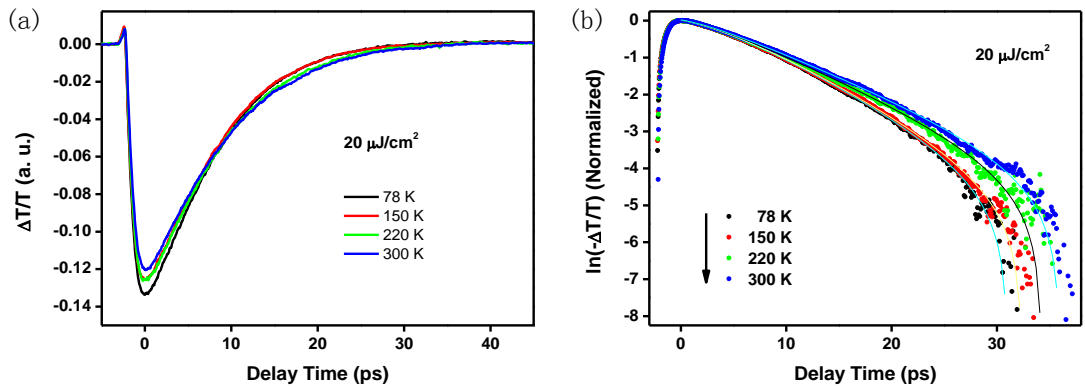

Figure S3. (a) Temperature dependence of transient THz responses of $Cd_3As_2$ excited at 400 nm with pump fluence of 20 $\mu J/cm^2$, and (b) the natural logarithmic plot of normalized transient THz responses, the solid lines are bi-exponential function fittings.

As the measurements are carried out with 400 nm pump excitation and 20 $\mu J/cm^2$ fluence, the $|dT/T|_{t=0}$ slightly decreases with increasing lattice temperature as shown in Figure S3(a), because of the decrease difference between the electron temperature and lattice temperature as discussed in the main text. Figure S3(b) shows the slowed down of the excited hot carriers relaxation with the increasing lattice temperature, and this behavior is more obvious at lower fluence (Figure 3(a) and 3(b)). The relaxation dynamics over the entire temperature range are well fitted using a biexponential function and the respective decay times are presented in Figure 3(c), from which the electron-optical phonon scattering process is found enhanced with the temperature while the subsequent hot phonon relaxation is slowed down because of the bottleneck effect of hot phonon cooling.

## S4. Fitting of THz conductivity by using Drude model

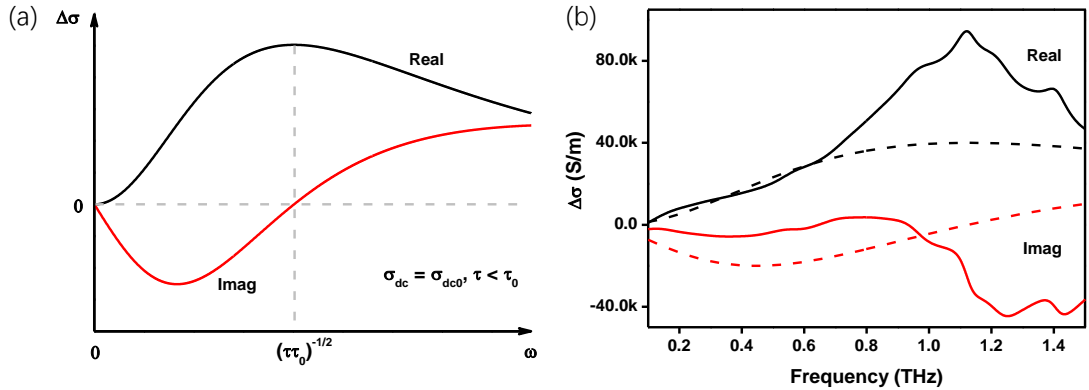

Figure S4. (a) The trends of Drude-like photoinduced THz conductivity $\Delta\sigma$ as $\sigma_{dc} = \sigma_{dc0}$ and $\tau < \tau_0$, (b) the measured THz conductivity $\Delta\sigma$ (solid lines) and the Drude fitting results (dash lines).

The photoinduced THz conductivity $\Delta\sigma = \sigma - \sigma_0$ can be fitted by using Drude model, where $\sigma = \sigma_{dc}/(1-i\omega\tau)$ is the conductivity with the pump excitation, $\tau$ is the mean scattering time of the carrier and $\sigma_{dc}$ is the steady-state conductivity, and $\sigma_0 = \sigma_{dc0}/(1-i\omega\tau_0)$ is the conductivity without the pump. Because the $\Delta\sigma$ at the lowest frequency of the spectrum is close to zero, this indicates $\sigma_{dc} \approx \sigma_{dc0}$. Therefore, the dominant parameters of $\Delta\sigma$ are $\tau$ and $\tau_0$, and the steady-state conductivity only influences the amplitude. Furthermore, the photoexcitation induces more carriers and the carriers scattering becomes more frequent and hence $\tau < \tau_0$, otherwise we will get a negative real part of $\Delta\sigma$, which conflicts our observations. This results to a $\Delta\sigma$ curves as shown in Figure S4(a). The real part is positive with a maximum value at $\omega = (\tau\tau_0)^{-1/2}$ and exhibited a sub-linear behavior close to this value while the imaginary part is negative as $\omega < (\tau\tau_0)^{-1/2}$ and then became positive for higher frequency. The minimum value of the $\Delta\sigma$ for imaginary part appeared at lower frequency than the maximum value of Δσ for real part. However, for the Δσ we measured, the real part approaches to its maximum value from the lower frequency superlinearly, and the imaginary part minimum is at higher frequency compared to the real part maximum. These differences make the Drude fitting model impossible as shown in Figure S4(b), where we use $\sigma_{dc} = \sigma_{dc0} = 120000$ S/m, $\tau = 0.1$ ps and $\tau_0 = 0.2$ ps to do the fitting. We attribute the differences to the phonons absorption.